# Development of local plasticity around voids during tensile deformation

Yi Guo[a*], Cui Zong[b], T. Ben Britton[a]

[a] Department of Materials, Royal School of Mines, Imperial College London, London SW7 2AZ, UK

[b] National key laboratory of advanced high temperature structural materials, Beijing Institute of Aeronautical Materials, Beijing, 100095, China

**Abstract**

Voids can limit the life of engineering components. This motivates us to understand local plasticity around voids in a nickel base superalloy combining experiments and simulations. Single crystal samples were deformed in tension with in-situ high angular resolution electron back scatter diffraction to probe the heterogeneous local stress field under load; the reference stress is informed by crystal plasticity finite element simulations. This information is used to understand the activation of plastic deformation around the void. Our investigation indicates that while the resolved shear stress would indicate slip activity on multiple slip systems, slip is reduced to specific systems due to image forces and forest hardening. This study rationalizes the observed development of plastic deformation around the void, aiding in our understanding of component failure and engineering design.

# 1. Introduction

The presence of voids or inclusions that may nucleate voids are common in engineering materials and yet a physical understanding to the process of void growth remain obscure, despite decades of effort studying the ductile fracture of materials relating to the growth and coalescence of voids. In a modern gas turbine engine, nickel base superalloy is widely used to endure the most demanding environment as turbine blade [1,2]. Shrinkage porosity is a common defect seen in cast superalloy [3,4] and it is known to affect the fatigue life by crack nucleation [5,6], creep cavitation [2], and creep crack formation [7].

The mechanical growth of void has been related to stress triaxiality through the Rice and Tracey model [8]. Stress triaxiality is defined as the ratio between hydrostatic stress and von Mises equivalent stress. Gurson and colleagues [9–11] developed a series of continuum constitutive theories linking hydrostatic stress to void growth during ductile fracture. The development of these theories has been facilitated by X-ray computed tomography (XCT) [12,13] which enables a more accurate account of some critical parameters, such as the evolution of void size and volume fraction during deformation. Though these theories have found wide adoption in solving engineering problems, the physical process of void growth was not clear. Progress has been made recently correlating void growth with local microstructures through advanced correlative approach combining XCT with scanning electron microscopy (SEM) and electron back scattered diffraction (EBSD) [14,15]. Noteworthy are some recent advancements in finite element simulation strategy that approaches the void growth and ductile fracture process in a physically based formulations [16,17] and multi-physics crystal plasticity framework of damage nucleation applicable to high rate deformation [18,19].

Molecular and dislocation simulations provide a mechanistic insight into void growth under multiaxial remote loading conditions [20–23]. These, together with some earlier work [24–26] indicate that mass transfer required for the expansion of void is primarily due to dislocation activity around the void surface. The dislocation emission based void growth models, schematically summarised in Fig. 1, generally involve prismatic dislocation loops [27–30], shear loops [27,31], antiparallel dislocations gliding on parallel slip planes [23,27,32], and antiparallel dislocations gliding on angled slip planes [33,34].

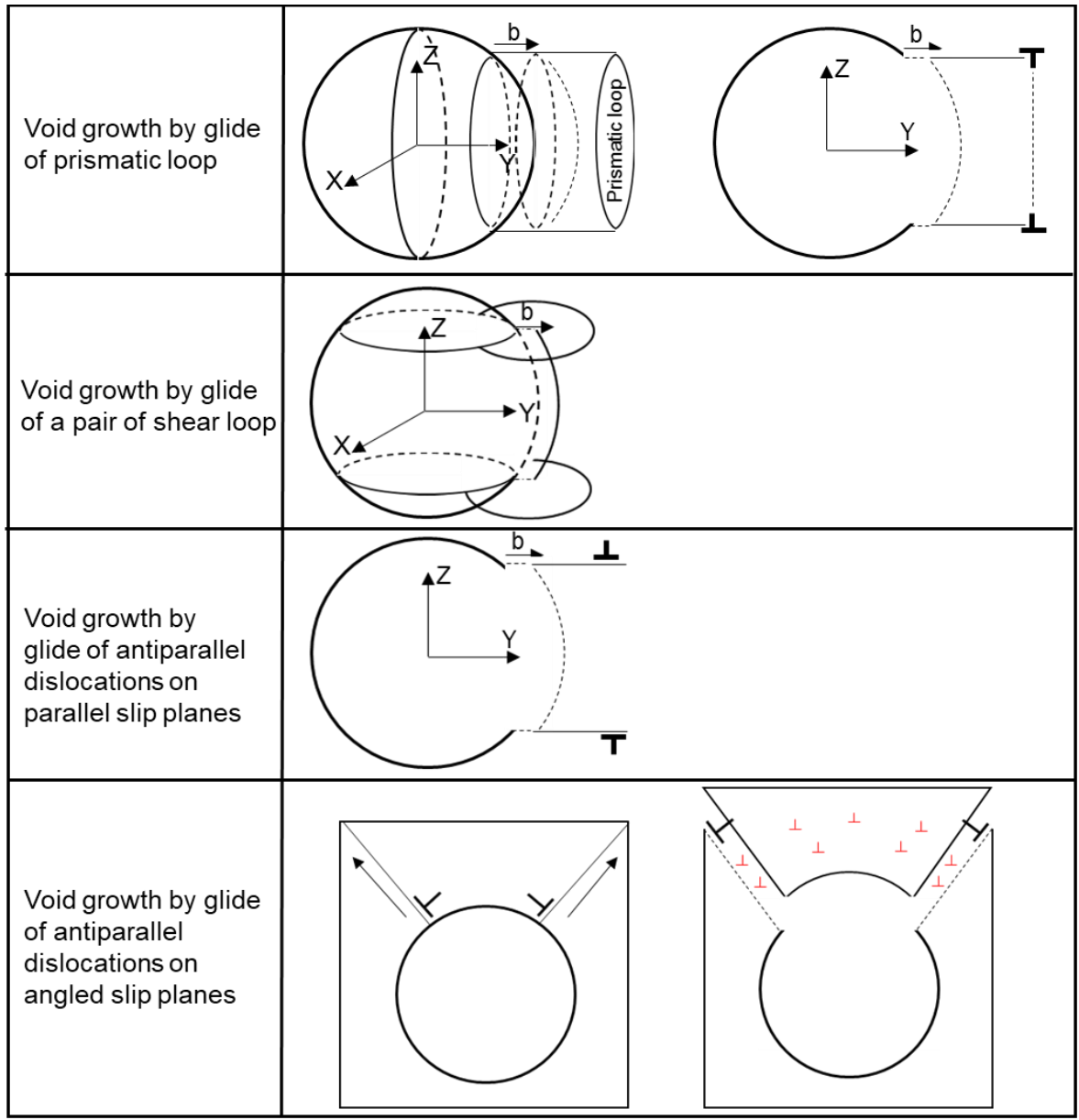

*Figure 1: Schematic representation of void growth mechanisms found in literature.*

The shear loop mechanism has been critically reviewed by Bulatov et al. [28]. The volume change of the void associated with a dislocation loop is [35]:

$$\delta V = \int_{Surface} (\boldsymbol{b} \cdot \boldsymbol{n})\, dA \quad (1)$$

Where **b** is Burgers vector of the dislocation loop, **n** is the unit normal to the surface *dA* bounded by the dislocation loop. This is to indicate that for shear loops the net mass transfer equals zero as *b* and *n* are perpendicular; i.e. When a full shear loop is punched out of the void surface the void would return to its original shape [28].

For void growth by glide of antiparallel dislocations on angled slip planes, shown in Fig. 1, this would create lattice curvature and need to be supported by the accumulation of geometrically necessary dislocations (GND).

These four mechanisms highlight how local plasticity can be involved in void growth and motivate the present study. Prior work has involved simulations and analytical studies and so in the present work we supplement this with direct measurements of the stress field and dislocation structures near voids during in-situ tensile testing. The stress field is evaluated using a combination of high resolution EBSD (HR-EBSD) [36–38] facilitated by crystal plasticity simulation [39]. Analysis of local plasticity is supported through careful consideration of the resolved shear stress of the 12 FCC slip systems and discussed in the context of image stress analysis [40,41]. While the investigations focus on the early stage of plastic deformation, the results suggest additional factors that affect void growth behaviour.

**2. Experiments and methods**

2.1. In-situ EBSD

The material used in this study is a second-generation low rhenium single crystal superalloy designated 'DD6'. A compositional comparison with other superalloy systems are made by Li. et.al [42] together with some general thermal and mechanical properties. The single crystal in this study was manufactured using directional solidification. From this, tensile test pieces were extracted using electro discharged machining. Before mechanical testing, the samples were ground and polished to a mirror finish (ending on 0.05 µm pH balanced colloidal silica) and the final surface damage was removed using a Gatan PECs ion milling system working at 8 keV with 4º gun tilt for 5 minutes. The tensile sample design and the in-situ tensile testing set up is shown in Fig. 2. The waisted sample gauge is used to focus the region of high strain towards the central region and to avoid probabilistic strain localization in an otherwise straight gauge. The sample was fixed onto a 70º pre-titled sample stage with stainless steel pins, along the Ø3.1 mm cut-outs toward the sample shoulder to align the sample with the loading direction, and tightened with tensile clamping grips. For enhanced conductivity the sample was connected to the SEM stage using carbon tape.

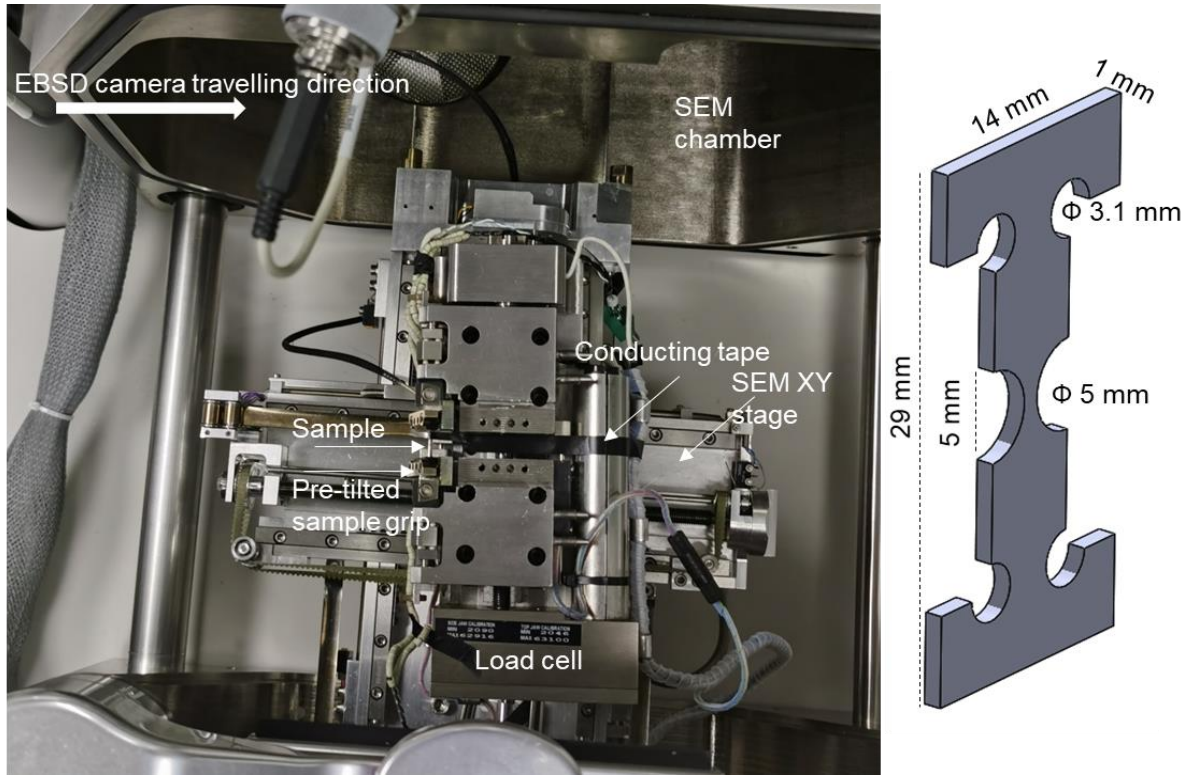

*Figure 2: In-situ testing setup and the sample design.*

The setup shown in Fig. 2 allows for a safe SEM/EBSD working distance of 18 mm and EBSD camera distance of 18 mm (for a phosphor which is 25.4 mm wide). The sample was deformed under displacement control at a rate of 0.1 mm/min and the displacement was paused for EBSD scans. A small region (~71 μm x 52 μm) surrounding the voids located in the middle of the gauge section was EBSD scanned at 30 kV and 14 nA with 0.4 μm step size and the diffraction patterns saved to 12-bit TIFF images for subsequent stress analysis.

The saved diffraction pattern images were used to analysis the rotations and strains present in the interaction volume that introduce shifts of the diffraction patterns [43]. The technique is known as high resolution EBSD (HR-EBSD) [36,37]. In principle, in the similarly orientated diffraction patterns (e.g. from one grain) one pattern is chosen as a reference pattern which is compared with other diffraction patterns to estimate displacement gradients. The displacement gradient tensors are then used to estimate strains and rotations which can, respectively, be related to stress, through anisotropic

Hooks law, and GND density, by linking rotation gradients to Nye's dislocation tensor [44,45]. It has been shown [38] that large lattice rotations, e.g. in a heavily deformed sample, affects the accuracy of strain measurement, in which case it is necessary to first rotate the intensity distribution within the test patterns close to the reference pattern and perform a second pass of pattern correlation to estimate strains. The validity of the HR-EBSD method has been confirmed by comparing with a synchrotron-based strain measurement technique [46,47].

2.2. Crystal plasticity finite element simulation

As mentioned above the stress estimated from EBSD are relative to a reference point whose stress are unknown. In this study, we use a lengthscale-dependent and physically based crystal plasticity simulation to extract the stress field at the reference point. The formulations of the simulation can be found in [39] and they are listed together with the simulation parameters in the Appendix. The simulation set up is shown in Fig. 3. The sample is deformed through the application of displacement to the right-hand surface, the x, y, z plane indicated on the left hand side of the sample shoulder in Fig. 2a is restricted movement along the corresponding directions, and the same is applied to the y and z plane of the right shoulder to avoid buckling at the sample gauge. To address the machine compliance, the model in Fig. 3 includes an isotropic elastic medium attached to the fixed end of the sample grip. A local region, of the same size of the EBSD scans, were assigned smaller element size as this is the primary region of interest, and the void is approximated as a 10 µm deep cylindrical void, as shown in Fig. 3b.

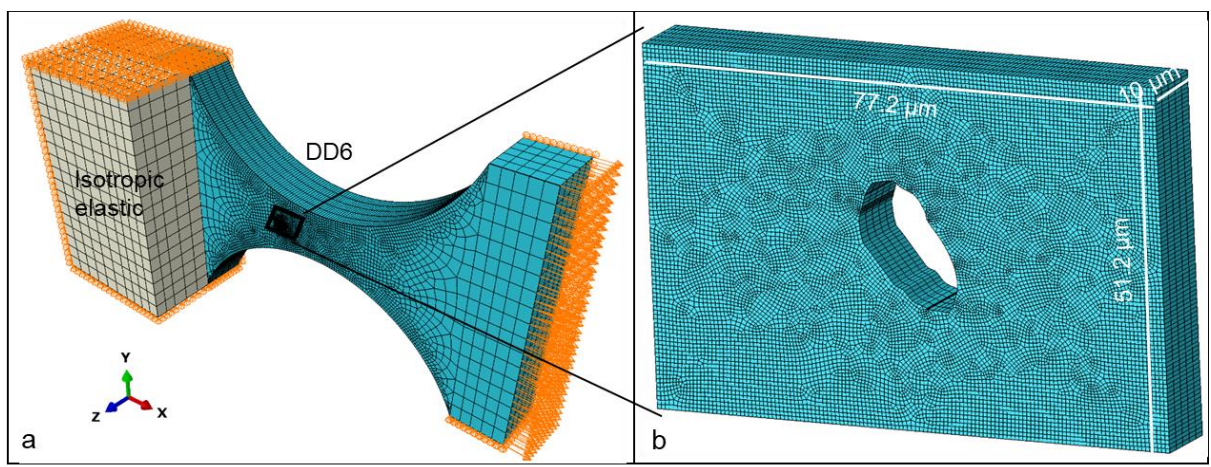

*Figure 3: Model set up for simulating the tensile property and the stress state at the HR-EBSD reference point.*

## 3. Results

The macroscopic load displacement curve and electron micrographs of the sample during a loading-unloading cycle are shown in Fig. 4. The loading is close to <001> direction, as revealed by the crystal orientation inset in Fig. 4-1, and the deformation features mainly the development of four slip systems, evidenced by the presence of the four sets of slip traces (indicated on Fig. 4-2) on the surface. The interactions between the slip bands likely result in the observed hardening.

EBSD scans were conducted at selected voids on the sample surface at different levels of deformations indicated on the load-displacement curve. The selected area imaging and EBSD scans resulted in a local build-up of carbon contamination (the black boxes). The timing of these observations include a pre-deformation scan (#1), a scan at yield point (#2), a scan at intermediate deformation (#3), a scan before fracture (#4, the displacement to fracture was determined by a separate experiment), and a final scan after unloading (#5).

At later stage of deformation, a tendency for sub-grain boundary formation can be seen towards the centre of the sample (Fig. 4 image #4). Between Fig.4 image 4 and 5 there are no visible changes of the surface deformation features after unloading.

The location of the higher magnification inserts in Fig. 4, which demonstrate the local area around one void, are indicated with a box in the low magnification images. These inserts reveal the development of local plasticity. Here localised dislocation plasticity has resulted in the formation of two slip traces from a location on the top of the void.

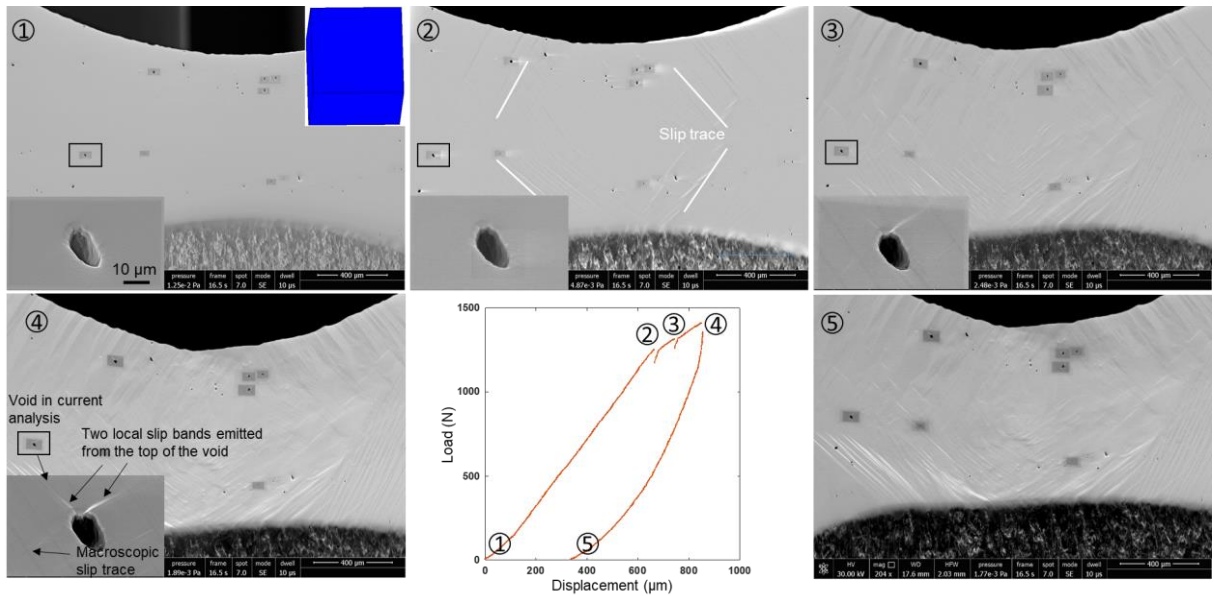

*Figure 4: The evolution of the surface deformation feature during a loading-unloading cycle. The inset shows the development of the local plasticity around a void (location indicated by the box).*

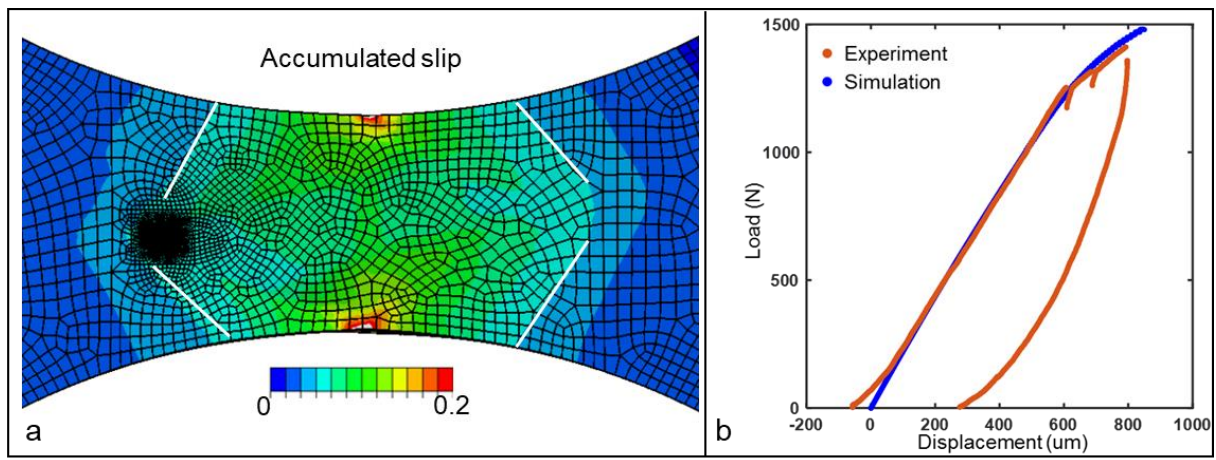

*Figure 5: The accumulated slip from simulation shows slip trace consistent with the experimental slip trace (a), and the comparison between experimental and simulated load-displacement curve shows good agreement(b).*

Results from the crystal plasticity model are shown in Fig. 5a, including the accumulated slip. The orientation of the bands with high strain gradient agree with those observed experimentally (indicated in Fig. 4-2). The statics and kinematics of the simulation show good agreement at the macroscopic scale, as there is agreement of the simulated and experimental load-displacement curve, including the transition from (macroscopic) elastic to plastic deformation as well as macroscopic hardening. This indicates that the simulations capture both the macroscopic mechanical response and the slip trace orientation and therefore the deformation within the experimentally measured crystal orientation. The void spacing was quantified from X-ray tomography study (Supplementary Fig. 5) and the voids with the nearest neighbour spacing smaller than 20 µm consists of 20% of the void population while the local plastic zone of the voids reaches 10-20 µm from the void surface. In addition, the void growth is limited to a couple of microns in this material (2.13 µm for the current void), therefore, the macroscopic mechanical response is less likely to be sensitive to the interactions between voids.

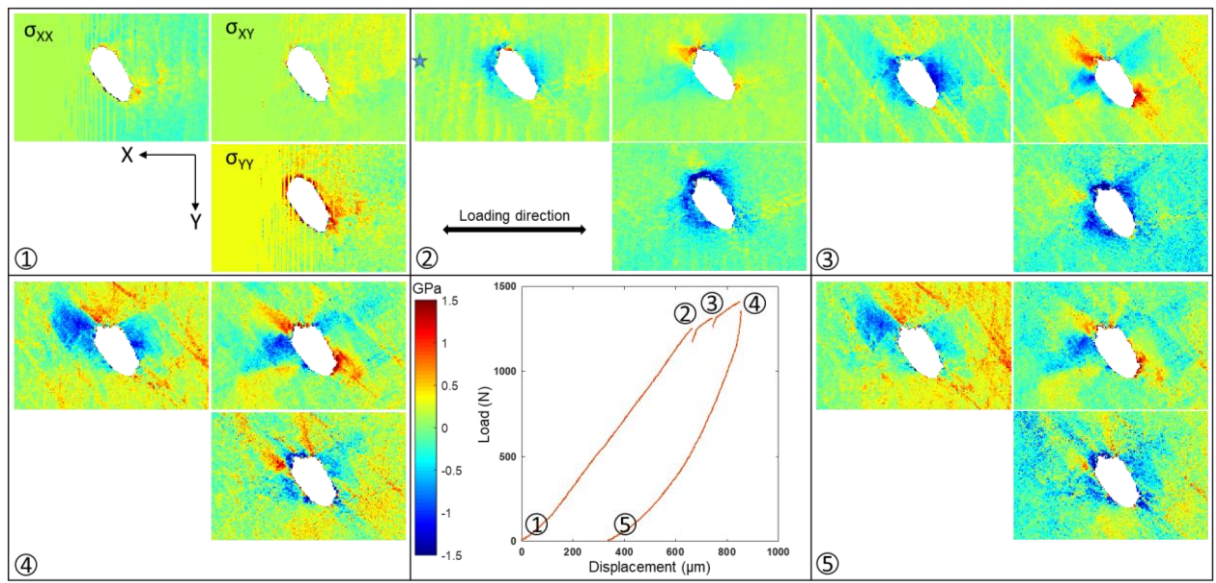

*Figure 6: The development of the local in-plane stress field at a void during a loading-unloading cycle.*

Voids are shown in Fig. 4 and they are stress concentrators during deformation, changing the local deformation conditions. The local stress field at one of the voids (location indicated in Fig.4-2) is demonstrated in Fig. 6, measured underload. The star in Fig.6-2 indicate the location of the reference point and is consistent in other deformed maps. In Fig. 6 each map presents the stress field around the void with respect to this reference point, i.e. these maps show variations in the Type III stress distributions.

The initial state of the sample reveals a heterogenous residual stress field, and the heterogenous nature of the stress field is homogenised around the yield point as indicated by Fig.6-2.

Comparison of the stress fields under load vs unloading indicates that the variations in these stress fields, from the yield point, persist until after unloading. For example, the $\sigma_{XX}$ stress is compressive to the left and right side of the void and tensile at the top and bottom. After unloading, there is no significant redistribution of the stress field, but the amplitude is reduced.

To address the reference stress issue, the crystal plasticity simulation is used to shift the stress tensor across the map according to the principle of superposition (i.e. add the simulated stress field at the reference point to other points in the experimentally measured stress maps). This follows the approach used previously (e.g. [48]) and this avoids use of a second experimental technique (e.g. synchrotron diffraction [49]). As a note, alternatively the simulation field could be 're-referenced' to reveal only the type III stress variations, but as we will see later the absolute value of the stress field is useful for mechanistic analysis.

To confirm the validity of this approach, the 're-referenced' experimental fields are compared with the simulation fields in Fig. 7 and good agreement for the spatial variations are shown. In these figures, the stress tensor at the reference point has a dominant $\sigma_{XX}$ component while the other components are negligible. This reference stress state contribution is likely because the void is located close to the central axis of the specimen and the reference point was chosen far away from the void and therefore was less affected by the local stress field developed around the void. The reference shifted HR-EBSD stress field is shown in Fig. 7c and the magnitude of the reference stress state is sufficient to render the average stress state of the map having a dominant $\sigma_{XX}$ tensile stress field, with strong variations close to the void.

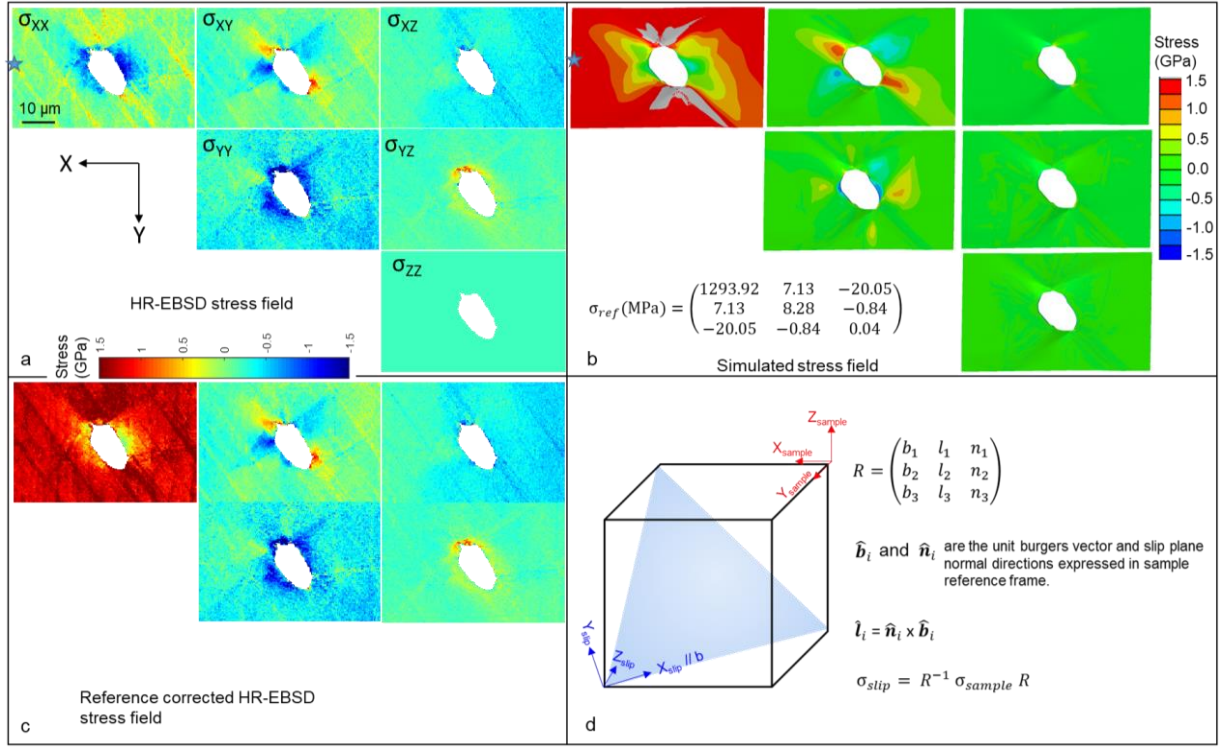

*Figure 7: Full stress field around the void, as measured by HR-EBSD (a) and simulated using crystal plasticity method (b) where the stress tensor at the reference point is demonstrated. The reference shifted HR-EBSD stress field is shown in (c). Two sets of reference frame are defined in (d) together with the method to rotate from sample reference frame to slip reference frame.*

The development of local plasticity around the void can be seen from Fig. 4. Slip trace is seen at the intermediate deformation level (scan 3) without further development of other slip systems between scan 3 and close to fracture (scan 4).

To further understand the local plasticity the resolved shear stress along each of the 12 FCC slip systems are investigated. Here, we use the reference shifted HR-EBSD stress field at scan 3 and scan 4 and assume that these would shed light on the lack of further development of local plasticity around the void between the two deformation levels. As the resolved shear stress is the major factor driving dislocation motion, it is of interest to investigate this for each of the 12 FCC slip systems. This is done by rotating the reference shifted HR-EBSD stress field from the sample reference frame to a reference frame on slip systems defined as X axis parallel to the Burgers vector direction and Z axis perpendicular to the slip plane. The method is schematically shown in Fig. 7d. The $\sigma_{XZ}^{slip}$ component of the rotated stress tensor represent the resolved shear stress of the corresponding slip system. The results are shown in Fig. 8. (Note in Fig. 7 and 8 we only show the stress fields at scan 3 for simplicity, those at scan 4 are shown in Supplementary Figures).

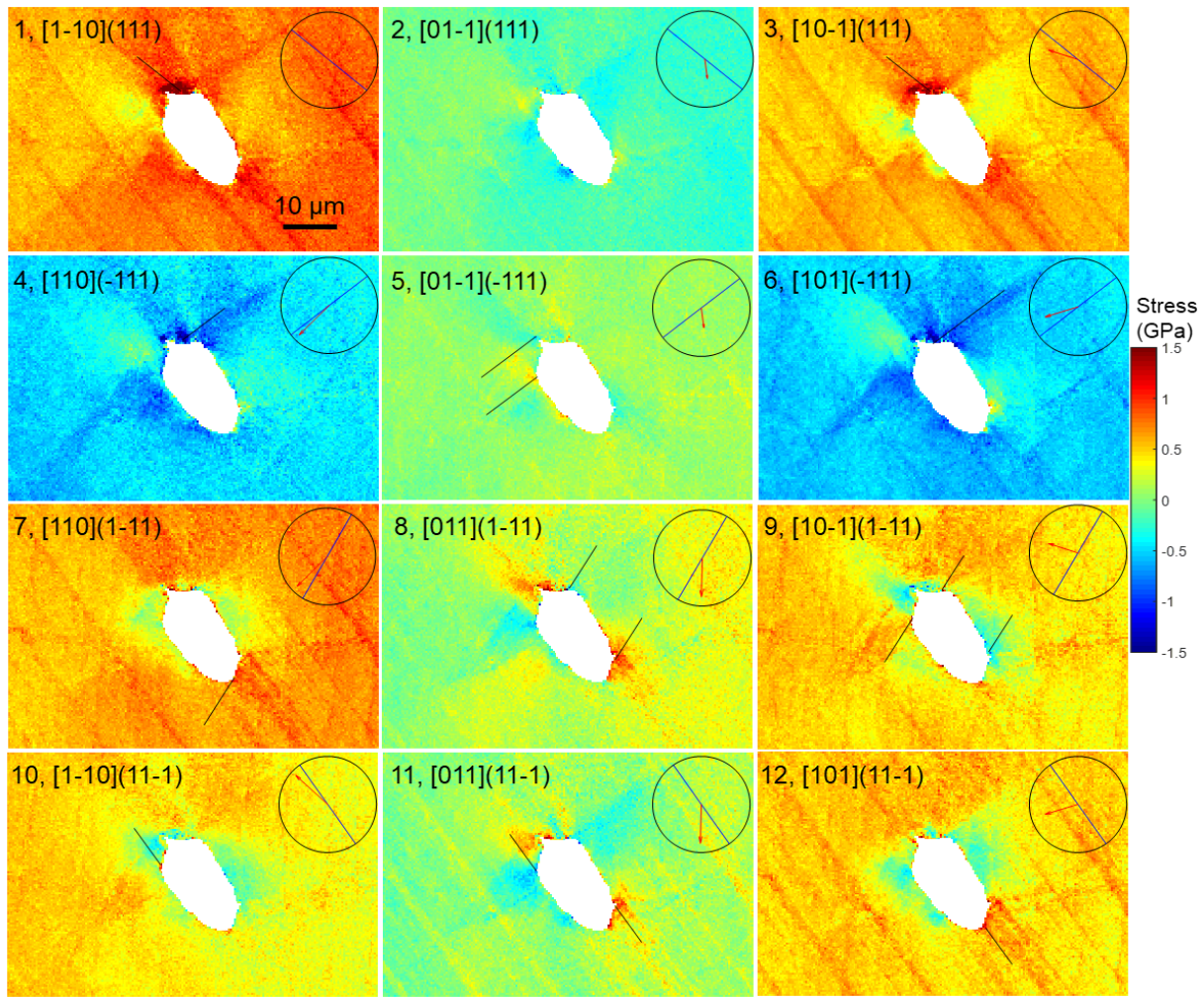

*Figure 8. The resolved shear stress maps of the 12 FCC slip systems. The clock diagrams indicate the slip trace, represented by the line going through the diameter of the circle, and the in-plane Burgers vector direction, represented by the arrow and the length of this arrow indicates how 'in-plane' the Burgers vector is.*

In Fig. 8 the slip system is labelled on the corresponding resolved shear stress map together with a clock chart indicating the corresponding slip trace (diametric line) and the in-plane Burgers vector direction (arrow). The length of the arrow represents the magnitude of the out-of-plane Burgers vector component, short arrow indicates Burgers vector pointing in or out of plane. The location of possible slip trace formation can be taken as the location where there is high magnitude of positive (along Burgers vector direction) or negative resolved shear stress field (against Burgers vector direction). The direction of resolved shear stress need to be directed away from the void surface so that dislocations can glide away to facilitate void growth. Based on this principle, the locations around the void where slip trace formation is possible are labelled with solid lines in Fig. 8. All the slip systems can potentially be initiated from multiple locations at the void surface, except slip system 2 (where the sign of the stress field does not align with the Burgers vector to result in slip away from the void).

Despite these apparently favourable stress conditions, Fig. 4 shows that only 2 slip systems are observed. These align with slip on the (111) and ($\bar{1}11$) planes. In the following section, we will explore why only these two systems were activated.

## 4. Discussion

Measurement of the stress state and observation of dislocation plasticity around the void motives us to test mechanism driven hypotheses for slip activity near the void. The size of voids in an engineering alloy can vary from sub-micron scale to micrometer scale and beyond, depending on the alloy system, thermal/mechanical history, and the level of plastic deformation. The variation of void size may results in differing growth rate due to strain gradient [50]. Previous studies have pointed out that the size effect is more prominent for sub-micron voids and its influence on micron voids is weak [51,52]. Note that this conclusion is valid for high triaxiality regime, i.e. around 4.5, as has been indicated in [53]. For the current material, the porosity is ~0.15 vol% and the void size ranges from 5 μm to 40 μm as have been confirmed by an X-ray tomography study (image supplied in Supplementary). This is typical of the void sizes of the casting superalloy that has been confirmed in literature [54,55]. The stress triaxiality of the current notch design (i.e. 2.5 mm notch radius, 1mm$^2$ cross section) falls below 1, as can be seen from literature data [56–58] and estimated using a Bridgman equation [59]. In this triaxiality regime, the void size effect is insignificant [53]. From these points of view, it is anticipated that the void studied in this investigation is representative of the void growth behaviour in this material.

Usually, slip is driven by the resolved shear stress on the slip system exceeding a critical value. The fields in Fig. 8 would indicate that significant slip should occur near the void, either through source nucleation or movement of existing dislocations.

However these stresses can be shielded and reduced, especially when dislocations are near a free surface due to the additional stress field called image stress [40,41]. The origin of image stress is schematically represented in Fig. 9, where the screw dislocation, represented by a Volterra cut, close to an interface would exert a shear stress, $\sigma_{XZ}$ and $\sigma_{YZ}$, on the interface under the designated reference frame [40]. The general equilibrium and compatibility requirement at the interface [40] (note the schematic only represent an equilibrium condition) requires an equal but opposite stress at the interface amounting to placing an 'image' dislocation of opposite sign at the other side of the interface. Dislocations may be attracted to or repelled from the interface depending on the difference of shear modulus across the interface [40]. In the case of a void where the interface is a free surface, an attraction force is exerted on the dislocations.

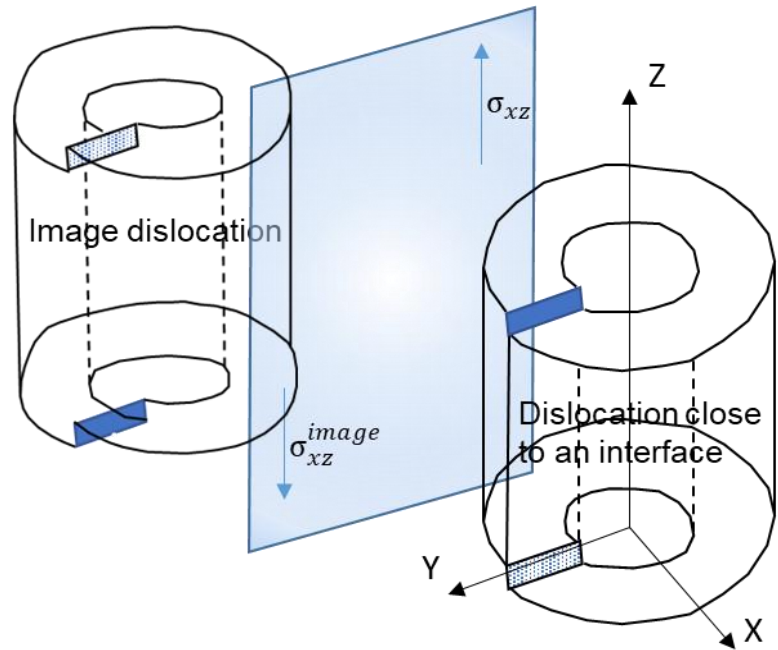

*Figure 9: Schematic illustration of the origin of image stress, reproduced from [40].*

Fig. 9 provides a context on the origin of image stresses using a simplified case where a screw dislocation (represented by a Volterra cut) lies close to an interface. For a straight edge dislocation close to a void surface, we adopt here an analytical image stress model proposed by Lubarda [33]:

$$\sigma^{image} = \begin{pmatrix} \frac{Gb_y}{2\pi(1-v)} \frac{x}{x^2-a^2} [\frac{a^2}{x^2} - (1-\frac{a^2}{x^2})^2] & -\frac{Gb_x}{2\pi(1-v)} \frac{x}{x^2-a^2} & 0 \\ & -\frac{Gb_y}{2\pi(1-v)} \frac{x}{x^2-a^2}(1+\frac{a^2}{x^2}-\frac{a^4}{x^4}) & -\frac{Gb_z}{2\pi} \frac{x}{x^2-a^2} \\ symmetric & & 0 \end{pmatrix} \quad (3)$$

Where *G* is shear modulus, $\nu$ is Poisson's ratio, *α* is void radius, *x* is the distance to the centre of the void, and $\boldsymbol{b} = (b_x\ b_y\ b_z)$ is the Burgers vector. The anisotropic shear modulus is used for the calculation and in the case of {111} slip in FCC crystals this is given by (from [60]):

$$G^{(111)} = \frac{3}{1+2A} C_{44} \quad (4)$$

Where $A = \frac{2C_{44}}{C_{11}-C_{12}}$ is the Zener anisotropy ratio, $C_{11}$ $C_{12}$ and $C_{44}$ are the stiffness constants (the corresponding values of 252 GPa, 161 GPa, and 131 GPa are used in this study). The resolved image stress on the 12 FCC slip systems are obtained by the routine explained in Fig. 7d and the results are presented in Fig. 10.

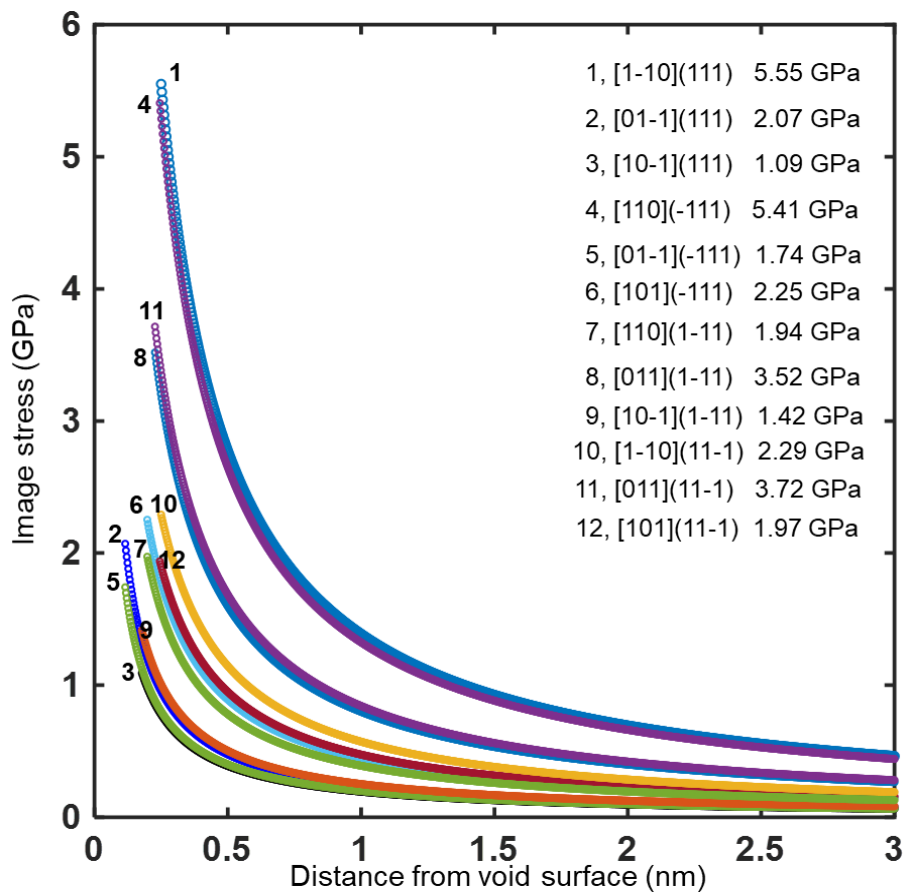

*Figure 10: Image stress of the 12 FCC slip systems. Labelled stress values are those at one full Burgers vector distance from the void surface.*

Notice the image stress vs distance plot are made from one Burgers vector away from the interface, and the analytical solution of (3) no longer holds due to distorted dislocation core when the distance to void surface is close to the dislocation core cut-off distance [33]. The 12 FCC slip systems together with the corresponding image stress at one Burgers vector distance are labelled in Fig. 10 and the stress magnitude represents a lower-bound resistance to dislocations gliding away from the void surface.

Now the resolved shear stresses on the 12 slip systems in Fig. 8 are revisited and considered in terms of the image stress field that tend to attract dislocations towards the free surface.

Comparing the stress fields in Fig 8. and the dislocation slip based void growth modes shown in Fig. 1, it can be seen that slip systems 5, 8, 9 have local stress fields that facilitate antiparallel dislocations glide on parallel slip planes. For instance, in the shear stress map for slip system 8 in Fig. 8, the negative stress field at the top right corner and the positive stress field at the bottom right corner of the void both tend to direct dislocations with opposite sign away from the void surface along the indicated slip traces. For slip system 7, 9, 10, 11, and 12, the corresponding shear field facilitates the formation of single or multiple local slip traces.

To provide quantitative assessment of slip activity, line scans are drawn along the indicated slip traces, starting from the pixel with the highest local stress concentrations (i.e. maximum driving force for dislocation motion), for the above two categories of slip systems, and the magnitude of the shear stresses are compared to the image stress of the corresponding slip system in Fig. 11. The comparison brings to our attention that the resolved shear stress level for most of the slip systems at scan 3 are on the order of 1 GPa lower than the image stress level. The gap is larger in some cases, notably slip systems 8 and 11. Though the resolved shear stresses show various level of increase due to continued macroscopic loading up to scan 4, the resolved shear stress for most of the slip systems are still below the image stress. The slip system 9-1 has resolved shear stress level that exceeds the image stress close to the free surface at scan 3 and at scan 4 the resolved shear stress for slip system 5-2 also rises to a value above the image stress level, however the corresponding slip traces (indicating that these systems activated) were not observed. A further exploration of the absence of these slip traces follows.

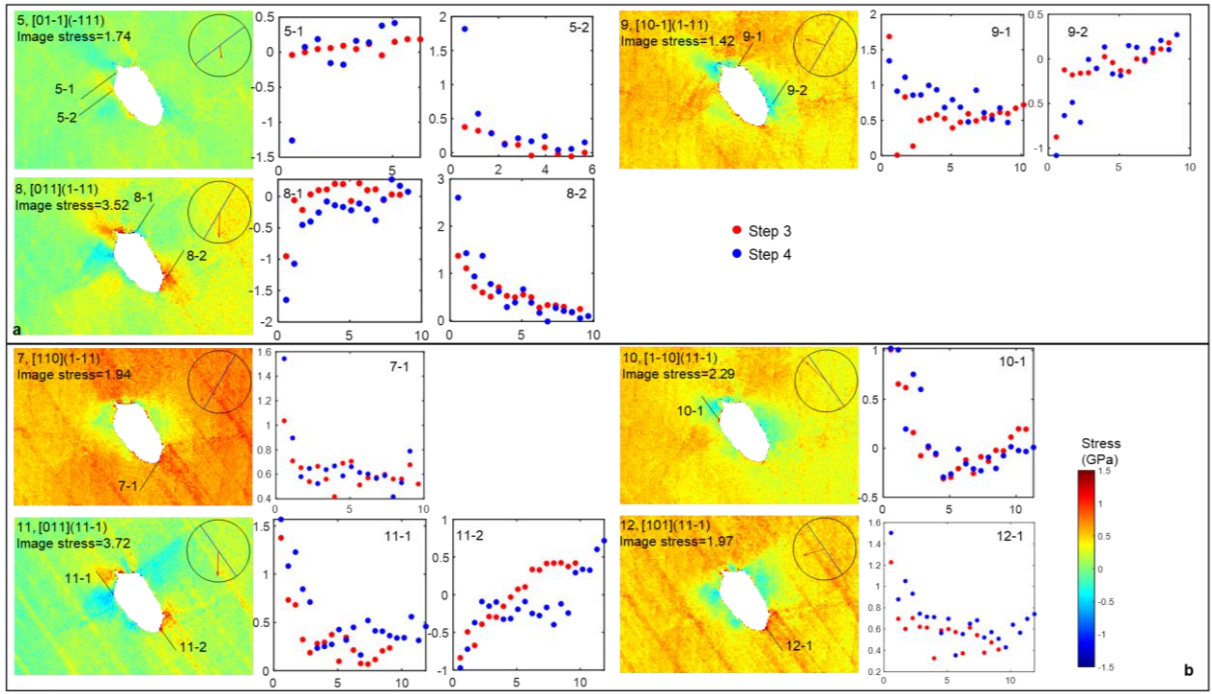

*Figure 11: Resolved shear stress maps facilitating the formation of antiparallel dislocations gliding on parallel slip planes (a) and single or multiple slip traces (b). The magnitude of the resolved shear stresses along the potential slip lines are compared with the image stresses.*

The shear stress field for slip systems 1, 3 and 4, 6 facilitates the formation of slip trace corroborating the observed slip trace (Fig. 4), a comparison with the image stresses are shown in Fig. 12a and b. These slip systems in general have higher resolved shear stresses compared to those in Fig. 11, however, much higher image stresses for slip systems 1 and 4 suggest that those two slip traces are less likely to be the activated slip trace. The shear stress for slip system 3 is higher than the corresponding image stress, it is likely that this variant consists of one of the observed slip trace (Fig. 4) while slip trace and Schmid factor analysis wouldn't make this distinction between slip 1 and slip 3 (both give rise to the same slip trace and both have high Schmid factor: 0.43 and 0.40 respectively). The resolved shear stress for slip system 6 is lower than the image stress but note that this observation was made at an instance after the formation of this slip trace, it is expected that a certain proportion of the resolved shear stress to be relaxed. Therefore, compared to the more significant stress gap for slip 4, slip 6 is likely the observed slip variant. Notice that slip 3 and slip 6 forms an antiparallel dislocation pair gliding along slip planes at an angle. This configuration may grow the void according to the fourth mechanism in Fig. 1 and the observed GND densities in between those two slip planes (Fig. 12) seems to agree with this growth mechanism. The Accumulation of the GNDs around the void provides forest hardening which further reduce the probability for dislocations gliding away from the void surface. A Taylor hardening model [61], $\tau = 0.9bG\sqrt{\rho_{GND}}$, is used to obtain the stress due to forest hardening and the result is shown in Fig. 12c. There is raised GND density and therefore Taylor stress field at the nucleation site of slip 3 and 6, possibly due to the lattice continuity requirement to assist void growth. The indicated location, where the slip 9-1 (Fig. 11) shows resolved shear stress exceeding the image stress at scan 3 and that for 5-2 at scan 4, has an additional slip resistance on the order of 1 GPa and 0.3 GPa, respectively, imposed by forest hardening. This would bring the net driving force below the image stress. Notice that the forest hardening stress is generally below 1GPa at both deformation scan 3 and scan 4 (GND field at scan 4 is shown in supplementary), except for the location where the observed slip trace emitted from the void surface. For a significant part along the void perimeter the forest hardening stress is lower than the magnitude of the resolved shear stress for many slip systems shown in Fig. 11, therefore forest hardening itself is not sufficient to rationalize the lack of local plasticity around the void. Note that the GND densities are based on rotation gradients only, while although the elastic strain gradients may be higher underload, their magnitude are in general lower than the rotation gradients and only 3 (out of 9) in-plane elastic strain gradient terms can be obtained from a 2D experiment. Therefore, it is expected that including those 3 elastic gradient terms makes insignificant contribution to the resultant GND density distribution, as has been demonstrated by Wilkinson and Randman [62].

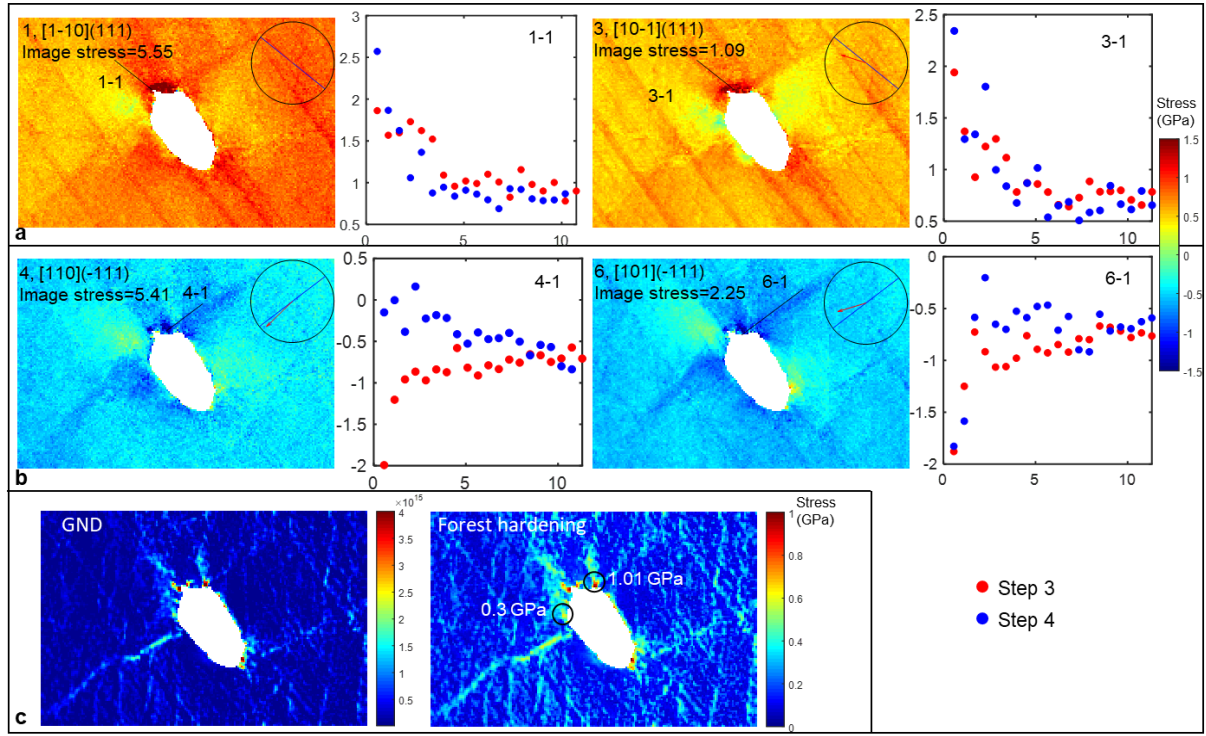

Figure 12: *Resolved shear stress maps facilitating the formation of slip traces corroborating the observed slip traces (a) and (b). The GND density distribution and the corresponding Taylor stress map are shown in (c). The x and y axis of the line profile plots represent the distance away from the void surface (µm) and the resolved shear stress (GPa) respectively.*

The void had possibly grown under mechanism 4 in Fig. 1. The equivalent diameter increased from ~16.6 µm (undeformed) to ~18.73 µm (before fracture), giving a growth ratio of ~1.13 at macroscopic strain of 0.085 (based on simulated stress strain curve, supplementary). At similar strain level, copper has shown a growth ratio of 1.29 [63] and aluminium alloy of 1.4 [64]. Additionally, significant and relatively uniform local plasticity can be seen around the void surface in these two materials [63,64].

The image stress can be considered as an energy barrier from moving dislocations away from the void surface. This stress is compared across various material systems in Fig. 14. For simplicity, Burgers vectors are assumed to be parallel to the horizontal axis and aligned with the centre of a 10 µm void. Therefore, the image stress is simply $\sigma_{xy}^{image} = -\frac{Gb}{2\pi(1-\nu)}\frac{x}{x^2-a^2}$ .

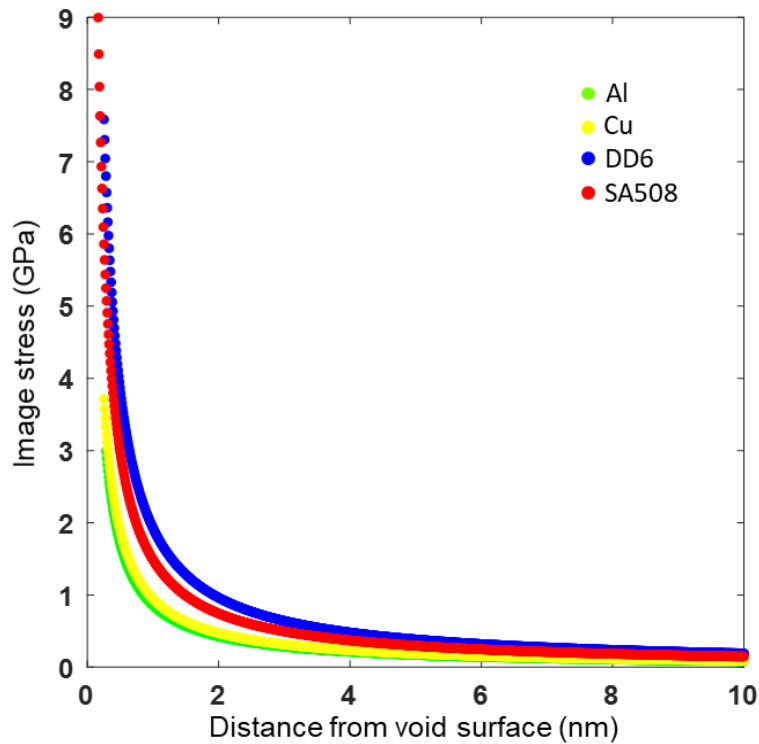

*Figure 13: Image stress vs distance from void surface plot across different material systems.*

It can be seen that the magnitude of the image stresses for Al and Cu do not pose an as significant barrier for void growth based upon dislocation glide, as compared to Ni superalloy (DD6), and this may be a factor influencing the well-developed local plasticity and void growth seen in these two materials [63,64]. Another material given as a comparison is a bainite steel (SA508) where the void imposes a higher image stress compared to DD6 yet demonstrate significant void growth [15,65]. Note that on the one hand, the SA508 steel has BCC crystal structure where slip on the {110}, {112}, and {123} planes [66] offer 36 slip systems compared to 12 in FCC Ni. On the other hand, the significant void growth in SA508 is typically either found in front of a crack [15] or after the UTS and thus during necking [65]. In these cases the local stress state may raise above the image stress for more slip variants and this can be combined with a consideration of the multiaxial stress state at the crack front and the necked region that may also facilitate dislocation emission by modifying the distribution of the local stress state [23] and this may result in more easy activation of a range of slip systems at the void surface, thus homogenising plasticity assisted void growth. This point can be directly seen from Fig. 6 of reference [21] where the dislocation distribution at void surface against various loading conditions were compared. The current investigation is limited to sample surface so the effect of multiaxial loading cannot be assessed, but the methodology can be readily applied to 3D studies should an appropriate instrument (e.g. Synchrotron radiation) is available.

For DD6, significant void growth has been observed, but typically at high temperature, for example, in Fig. 6 of reference [67] where transition has been seen from a mixed ductile/cleavage fracture at 850°C to a full ductile fracture featuring void growth and coalescence at 1020°C. Note that at these high temperatures thermally assisted dislocation jump may help overcome the image stress as well as cross slip that facilitates prismatic loop formation. These are likely the factors contributing to the significant void growth seen at high temperature [67] while diffusion may not be a dominant mechanism given the time scale of the experiment in [67].

In the present work, Fig. 10 shows that that the image stress decrease quickly away from the free surface. All slip systems have image stress dropped below 500 MPa at 3 nm distance (i.e. ~90 atomic layers) away from the surface, and at this distance some slip systems have image stresses below 100 MPa. This is important to consider with regards to the measurement of GND densities with EBSD (though a different stress function is needed to properly analysis image stress of a flat surface). For an EBSD experiments, GND density is calculated using spatial gradients of the average lattice orientation of the interaction volume which extends ~20 nm from the surface [68] and therefore dislocations affected by image stress field consists of only a small portion. On the other hand, the narrow window from the void surface where the image stress is significant poses difficultly in the direct comparison with the current experimental measurements. While our study provides the first-order assessment and this suggests that more work using higher spatial resolution stress measurements will be valuable, despite the inherent challenges involved.

The image stress analysis presented in the current work shed light on dislocation mobility after nucleation. The nucleation of dislocations from void surface involves dislocation energy variations across slip systems, elastic anisotropy, dislocation core structure [69], as well as surface energy due to ledge formation [21,27]. Some detailed nucleation process can be found in [31,69]. The surface energy required to create a ledge from a smooth void surface may pose a barrier for dislocation nucleation hence restrict the local plasticity around the void, however, the native ledges or steps on the void surface are found to facilitate dislocation nucleation [21,70]. It has been suggested that the stress required to nucleate dislocations from void surface could be higher than the strength of the material [71], indeed, the critical stress to nucleate a dislocation from a 4 nm void in copper is on the order of 6 GPa [31] which is far higher than the strength of copper. This can only be practically achieved by high rate or low temperature deformation [71]. However, these molecular dynamics studies concern voids in nano-metre scale where the void size is on the order of the mean dislocation spacing. Under this condition, the dislocation nucleation consists a significant part of void growth due to the starvation of dislocation sources. For a micron-sized void, it is arguable that sufficient dislocations are already present along the void perimeter. Therefore, the growth of micron-voids may primarily involve dislocation gliding rather than nucleation.

## 5. Conclusion

In this study the stress field around a void in a nickel base superalloy was measured in-situ using HR-EBSD while the specimen was held under load. The stress state at the HR-EBSD reference point was estimated by crystal plasticity simulation, enabling an adjustment of the full field map to enable the assessment of the development of local plasticity at the void surface.

Notable observations are summarized below:

1. Assessment of stress while the sample is still under load is necessary in cases where the magnitude of the stress is important, since there are load drops after unloading (or an unloading model must be used). In this example, the distribution of stress field seems to be less affected by unloading.
2. For the material in this study, there was limited local plasticity development at the void surface throughout the deformation. This is evidenced by the observation that only two slip traces emitted from the top surface of the void.
3. Resolved shear stress analysis found multiple favourable locations around the void surface where slip trace formation is possible. However, preliminary investigations indicated at these locations the resolved shear stress fell below the image stress and thus it is assumed that this would provide additional barrier for dislocation motion hence inhibiting void growth.

4. GND accumulation at the void surface provide forest hardening and further resist dislocations gliding away from the void surface, but forest hardening alone cannot rationalize the lack of plasticity at void surface.
5. The void growth in this study is due to antiparallel dislocations glide on non-parallel slip plane. This mechanism, however, accumulates GNDs in between the slip traces, which may limit the growth rate by local forest hardening.
6. Image stress can be used to study/predict the dislocation activity close to void surface. It may be used to pin down the specific slip variant that can be activated where the traditional Schmid factor and slip trace analysis fails.
7. The image stress decreases quickly away from the void surface and is a function of shear modulus. The rate of decay depends on Burgers vector orientation. In general, the image stress in Ni superalloy drop below 500 MPa at 3 nm distance away from the void surface.

**Author contributions**

YG conducted the experiment and simulation, performed the data analysis, and drafted the paper. CZ manufactured the material. TBB supervised the overall project.

**Acknowledgement**

TBB acknowledges funding of his research fellowship from the Royal Academy of Engineering. The authors would like to acknowledge the support from Beijing Institute of Aeronautical Materials (BIAM). The research was performed at the BIAM–Imperial Centre for Materials Characterisation, Processing and Modelling at Imperial College London. The microscope and loading frame used to conduct these experiments were supported through funding from Shell Global Solutions and provided as part of the Harvey Flower EM suite at Imperial. YG acknowledges Mr. Alex Bergsmo, Mr. Vassilios Karamitros, and Prof. Fionn Dunne for their helpful discussions on crystal plasticity simulation. We wish to acknowledge Mr. Daniel Sykes and Mr. Zhuocheng Xu for help with the XCT experiment and the support of the Henry Royce Institute for enabling access to Zeiss Versa 520 facilities at Royce@Manchester; EPSRC Grant Number EP/R00661X/1.

**Appendix:** Crystal plasticity finite element formulation

The total deformation gradient $\boldsymbol{F}$ is decomposed into an elastic ($\boldsymbol{F}^e$) and a plastic deformation gradient tensor ($\boldsymbol{F}^e$) as

$$\boldsymbol{F} = \boldsymbol{F}^e \boldsymbol{F}^p \quad (1)$$

The plastic velocity gradient ($\boldsymbol{L}^p$) is determined from dislocation slip rate ($\dot{\gamma}^\alpha$) on the $\alpha^{th}$ slip system by

$$\boldsymbol{L}^p = \dot{\boldsymbol{F}}^p \boldsymbol{F}^{p-1} = \sum_{\alpha=1}^{12} \dot{\gamma}^\alpha \boldsymbol{s}^\alpha \otimes \boldsymbol{n}^\alpha \quad (2)$$

Where $\boldsymbol{s}^\alpha$ and $\boldsymbol{n}^\alpha$ are the slip shear (Burgers vector) direction and slip plane normal direction of the $\alpha^{th}$ slip system respectively. The slip rate ($\dot{\gamma}^\alpha$) is given by

$$\dot{\gamma}^\alpha = \rho_m b^{\alpha 2} v \exp\left(-\frac{\Delta F}{kT}\right) \sinh\left(\frac{(\tau^\alpha - \tau_c^\alpha)\Delta V^\alpha}{kT}\right) \quad (3)$$

Where $\rho_m$ is the mobile dislocation density, $b^\alpha$ is the Burgers vector magnitude of the $\alpha^{th}$ slip system, $v$ is the jump frequency of dislocations to overcome obstacles, $k$ is the Boltzman's constant, $T$ the thermal dynamic temperature. $\tau^\alpha$ is the resolved shear stress on the $\alpha^{th}$ slip system and mobilizes dislocations once it exceeds a critical value, $\tau_c^\alpha$. The rate sensitivity is controlled by the activation energy, $\Delta F$, and the activation volume, $\Delta V^\alpha$.

The Taylor hardening model is employed to account for the hardening of slip systems due to the accumulation of dislocations. Therefore, the evolution of the critical resolved shear stress ($\tau_c^\alpha$) is given by

$$\tau_c^\alpha = \tau_0^\alpha + Gb\sqrt{\rho_{SSD} + \rho_{GND}} \quad (4)$$

Where $\tau_0^\alpha$ is the initial lattice friction of the slip system $\alpha$, $G$ is the shear modulus, and $\rho_{SSD}$ is the statistically stored dislocation density whose evolution is a linear function of the effective plastic strain rate ($\dot{p}$)

$$\dot{\rho}_{SSD} = \lambda\dot{p} \quad (5)$$

in which $\lambda$ is an isotropic hardening coefficient.

The effective plastic strain rate ($\dot{p}$) is linked to the plastic deformation rate ($\boldsymbol{D}^p$) as

$$\dot{p} = (\tfrac{2}{3}\boldsymbol{D}^p : \boldsymbol{D}^p)^{1/2} \quad (6)$$

Where $\boldsymbol{D}^p$ is obtained by

$$\boldsymbol{D}^p = \text{sym}(\boldsymbol{L}^p) \quad (7)$$

The GND density is determined by relating the Nye dislocation tensor ($\boldsymbol{\Lambda}$) with lattice curvature

$$\boldsymbol{\Lambda} = \text{curl}(\boldsymbol{F}^p) = \sum_{\alpha=1}^{12} \rho_s^\alpha \boldsymbol{b}^\alpha \otimes \boldsymbol{s}^\alpha + \rho_{en}^\alpha \boldsymbol{b}^\alpha \otimes \boldsymbol{n}^\alpha + \rho_{em}^\alpha \boldsymbol{b}^\alpha \otimes \boldsymbol{m}^\alpha \quad (8)$$

Where the $\rho_s^\alpha$, $\rho_{en}^\alpha$, and $\rho_{em}^\alpha$ represent, respectively, the screw dislocation along the shear direction ($\boldsymbol{s}^\alpha$), edge dislocation along slip plane normal ($\boldsymbol{n}^\alpha$), and edge dislocation along the $\boldsymbol{m}^\alpha$ direction ($= \boldsymbol{s} \times \boldsymbol{n}$) of the $\alpha^{th}$ slip system. The plastic deformation was homogenized into the slip of 12 FCC slip systems and equation 8 is solved using a L2-norm minimization. GND density is obtained by

$$\rho_{GND} = \sqrt{\sum_{\alpha=1}^{12}((\rho_s^\alpha)^2 + (\rho_{en}^\alpha)^2 + (\rho_{em}^\alpha)^2)} \quad (9)$$

The above formulations are implemented through a user-material (UMAT) subroutine using ABAQUS standard analysis. The simulation parameters are based on reference [72] and chosen to minimize rate sensitivity. They are summarized in Appendix Table 1 together with the elastic constants:

*Appendix Table 1: Single crystal elastic constants and simulation parameters.*

| $\rho$ (μm$^2$) | $b$ (μm) | $\nu$ (s$^{-1}$) | $\Delta H$ (J) | $T$ (K) | $\tau_c$ (MPa) | $\lambda$ (m$^{-2}$) | $\Delta V$ |
|---|---|---|---|---|---|---|---|
| 1x10$^{10}$ | 2.61 x 10$^{-4}$ | 1.0 x 10$^{11}$ | 3.456 x 10$^{-20}$ | 293 | 450 | 3.5 x 10$^{14}$ | 11.6b$^3$ |

Elastic constants: $C_{11}$=252 GPa, $C_{12}$=161 GPa, $C_{44}$=131 GPa

The simulated load-displacement curve was brought into agreement with the experimental curve by adjusting the hardening parameter ($\lambda$) and the critical resolved shear stress ($\tau_c$). The model uses three-dimensional, 20-noded quadratic hexahedral elements with reduced integration (C3D20R, 6 DoF at each node). The simulation (251530 elements with 1090144 nodes) completed in 239 hours on a 44 core CPU, 128 GB Ram workstation.

**Data Availability**

The stress, strain, and GND maps, as measured using HR-EBSD in sample reference frame, have been deposited at Zenodo under the same title as this publication (http://doi.org/10.5281/zenodo.4672623).

**References:**

[1]   B.H. Kear, E.R. Thompson, Aircraft gas turbine materials and processes, Science (80-. ). 208 (1980) 847–856.

[2]   R.C. Reed, The superalloys: fundamentals and applications, Cambridge University Press, 2008.

[3]   M.R. Orlov, Pore formation in single-crystal turbine rotor blades during directional solidification, Russ. Metall. 2008 (2008) 56–60.

[4]   E. Plancher, P. Gravier, E. Chauvet, J. Blandin, E. Boller, G. Martin, L. Salvo, P. Lhuissier, Tracking pores during solidification of a Ni-based superalloy using 4D synchrotron microtomography, Acta Mater. 181 (2019) 1–9.

[5]   K. Prasad, R. Sarkar, K. Gopinath, Role of shrinkage pores, carbides on cyclic deformation behaviour of conventionally cast nickel base superalloy CM247LC at 870oC, Mater. Sci. Eng. A. 654 (2016) 381–389.

[6]   L. Kunz, P. Lukas, R. Konecna, High-cycle fatigue of Ni-base superalloy inconel 713LC, Int. J. Fatigue. 32 (2010) 908–913.

[7]   J.B. Le Graverend, J. Adrien, J. Cormier, Ex-situ X-ray tomography characterization of porosity during high-temperature creep in a Ni-based single-crystal superalloy: toward understanding what is damage, Mater. Sci. Eng. A. 695 (2017) 367–378.

[8]   J.R. Rice, D.M. Tracey, On the ductile enlargement of voids in triaxiality stress fields, J. Mech. Phys. Solids. 17 (1969) 201–217.

[9]   A.L. Gurson, Continuum Theory of Ductile Rupture by Void Nucleation and Growth: Part I-Yield Criteria and Flow Rules for Porous Ductile Media, J. Eng. Mater. Technol. 99 (1977) 2–15.

[10]  V. Tvergaard, Influence of voids on shear band instabilities under plane strain conditions, Int. J. Fract. 17 (1981) 389–407.

[11]  V. Tvergaard, A. Needleman, Analysis of the cup-cone fracture in a round tensile bar, Acta Metall. 32 (1984) 157–169. doi:10.1016/0001-6160(84)90213-X.

[12]  E. Maire, P.J. Withers, Quantitative X-ray tomography, Int. Mater. Rev. 59 (2014) 1–

43.

[13]  L. Babaut, E. Maire, J.Y. Buffiere, R. Fougeres, Characterization by X-ray computed tomography of decohesion, porosity growth and coalescence in model metal matrix composites, Acta Mater. 49 (2001) 2055–2063.

[14]  H. Toda, A. Takijiri, M. Azuma, S. Yabu, K. Hayashi, D. Seo, M. Kobayashi, K. Hirayama, A. Takeuchi, K. Uesugi, Damage micromechanisms in dual-phase steel investigated with combined phase-and absorption-contrast tomography, Acta Mater. 126 (2017) 401–412.

[15]  M. Daly, T.L. Burnett, E.J. Pickering, O.C.G. Tuck, F. Leonard, R. Kelley, P.J. Withers, A.H. Sherry, A multi-scale correlative investigation of ductile fracture, Acta Mater. 130 (2017) 56–68.

[16]  C. Ling, S. Forest, J. Besson, B. Tanguy, F. Latourte, A reduced micromorphic single crystal plasticity model at finite deformations, Int. J. Solids Struct. 134 (2018) 43–69.

[17]  X. Han, J. Besson, S. Forest, B. Tanguy, S. Bugat, A yield function for single crystals containing voids, Int. J. Solids Struct. 50 (2013) 2115–2131.

[18]  D.J. Luscher, J.R. Mayeur, H.M. Mourad, A. Hunter, M.A. Kenamond, Coupling continuum dislocation transport with crystal plasticity for application to shock loading conditions, Int. J. Plast. 76 (2016) 111–129.

[19]  T. Nguyen, D.J. Luscher, J.W. Wilkerson, A dislocation based crystal plasticity framework for dynamic ductile failure of single crystals, J. Mech. Phys. Solids. 108 (2017) 1–29.

[20]  H.J. Chang, J. Segurado, J. Llorca, Three-dimensional dislocation dynamics analysis of size effects on void growth, Scr. Mater. 95 (2015) 11–14.

[21]  H.-J. Chang, J. Segurado, O. Rodriguez de la Fuente, B.M. Pabon, J. Llorca, Molecular dynamics modeling and simulation of void growth in two dimensions, Model. Simul. Mater. Sci. Eng. 21 (2013) 1–17.

[22]  R.B. Sills, B.L. Boyce, Void growth by dislocation adsorption, Mater. Res. Lett. 8 (2020) 103–109.

[23]  J. Segurado, J. Llorca, Discrete dislocation dynamics analysis of the effect of lattice orientation on void growth in single crystals, Int. J. Plast. 26 (2010) 806–819.

[24]  M.F. Ashby, L. Johnson, On the generation of dislocations at misfitting particles in a ductile matrix, Philos. Mag. 20 (1969) 1009–1022.

[25] R.E. Rudd, J.F. Belak, Void nucleation and associated plasticity in dynamic fracture of polycrystalline copper: an atomistic simulation, Comput. Mater. Sci. 24 (2002) 148–153.

[26] J. Marian, J. Knap, M. Ortiz, Nanovoid Cavitation by Dislocation Emission in Aluminum, Phys. Rev. Lett. 93 (2004) 165503.

[27] V.A. Lubarda, M.S. Schneider, D.H. Kalantar, B.A. Remington, M.A. Meyers, Void growth by dislocation emission, Acta Mater. (2004). doi:10.1016/j.actamat.2003.11.022.

[28] V.. Bulatov, W.G. Wolfer, M. Kumar, Shear impossibility: comments on "Void growth by dislocation emission" and "Void growth in metals: atomistic calculations," Scr. Mater. 21 (2010) 2071–2088.

[29] T. Ohashi, Crystal plasticity analysis of dislocation emission from micro voids, Int. J. Plast. 21 (2005) 2071–2088.

[30] D.C. Ahn, P. Sofronis, R. Minich, On the micromechanics of void growth by prismatic-dislocation loop emission, J. Mech. Phys. Solids. 54 (2006) 735–755.

[31] S. Traiviratana, E.M. Bringa, D.J. Benson, M.A. Meyers, Void growth in metals: Atomistic calcualtions, Acta Mater. 56 (2008) 3874–3886.

[32] M.R. Gungor, D. Maroudas, Molecular-dynamics study of the mechanism and kinetics of void growth in ductile metallic thin films, Appl. Phys. Lett. 77 (2000) 343–345.

[33] V.A. Lubarda, Image force on a straight dislocation emitted from a cylindrical void, Int. J. Solids Struct. 48 (2011) 648–660.

[34] M. Huang, Z. Li, C. Wang, Discrete dislocation dynamics modelling of microvoid growth and its intrinsic mechanism in single crystals, Acta Mater. 55 (2007) 1378–1396.

[35] J.P. Hirth, J. Lothe, Theory of Dislocations, 2nd ed., Wiley, New York, 1982.

[36] A.J. Wilkinson, G. Meaden, D.J. Dingley, High-resolution elastic strain measurement from electron backscatter diffraction patterns: new level of sensitivity, Ultramicroscopy. 106 (2006) 307–313.

[37] T.B. Britton, A.J. Wilkinson, Measurement of residual elastic strain and lattice rotations with high resolution electron backscatter diffraction, Ultramicroscopy. 111 (2011) 1395–1404.

[38]   T.B. Britton, A.J. Wilkinson, High resolution electron backscatter diffraction measurements of elastic strain variations in the presence of large lattice roations, Ultramicroscopy. 114 (2012) 82–95.

[39]   F.P.E. Dunne, D. Rugg, A. Walker, Lengthscale-dependent, elastically anisotropic, physically-based hcp crystal plasticity: Appliation to cold-dwell fatigue in Ti alloys, Int. J. Plast. 23 (2007) 1061–1083.

[40]   J.P. Hirth, The influence of grain boundaries on mechanical properties, Metall. Trans. 3 (1972).

[41]   J.S. Koehler, On the Dislocation Theory of Plastic Deformation, Phys. Rev. 60 (1941) 397–410.

[42]   J.R. Li, Z.G. Zhong, D.Z. Tang, S.Z. Liu, P. Wei, P.Y. Wei, Z.T. Wu, D. Huang, M. Han, A low cost second generation single crystal superalloy DD6, in: Int. Symp. Superalloys, 2000: pp. 777–784.

[43]   S.I. Wright, M.M. Nowell, D.P. Field, A review of strain analysis using electron backscatter diffraction, Microsc. Microanal. 17 (2011) 316–329.

[44]   J.F. Nye, Some geometrical relations in dislocated crystals, Acta Met. 1 (1953) 153–162.

[45]   Y. Guo, D.M. Collins, E. Tarleton, F. Hofmann, A.J. Wilkinson, T.B. Britton, Dislocation density distribution at slip band-grain boudary intersections, Acta Mater. 182 (2020) 172–183.

[46]   Y. Guo, D.M. Collins, E. Tarleton, F. Hofmann, J. Tischler, W. Liu, R. Xu, A.J. Wilkinson, T.B. Britton, Measurements of stress fields near a grain boundary: Exploring blocked arrays of dislocations in 3D, Acta Mater. 96 (2015) 229–236.

[47]   S. Das, F. Hofmann, E. Tarleton, Consistent determination of geometrically necessary dislocation density from simulations and experiments, Int. J. Plast. 109 (2018) 18–42.

[48]   T. Zhang, D.M. Collins, F.P.E. Dunne, B.A. Shollock, Crystal plasticity and high-resolution electron backscatter diffraction analysis of full-field polycrystal Ni superalloy strains and rotations under thermal loading, Acta Mater. 80 (2014) 25–38.

[49]   B.C. Larson, W. Yang, G.E. Ice, J.Z. Tischler, Three-dimensional X-ray structural microscopy with submicrometre resolution, Nature. 415 (2002) 887–890.

[50]   N.A. Fleck, J.W. Hutchinson, Strain gradient plasticity, in: Adv. Appl. Mech., Academic Press, New York, 1997: pp. 295–361.

[51] P.F. Thomason, The Mechanics of Microvoid Nucleation and Growth in Ductile Metals, in: Ductile Fract. Met., Pergamon Press, Oxford, 1990: pp. 30–55.

[52] Y. Huang, H. Gao, W.D. Nix, J.W. Hutchinson, Mechanism-based strain gradient plasticity-II. Analysis, J. Mech. Phys. Solids. 48 (2000) 99–128.

[53] B. Liu, Y. Huang, K.C. Hwang, M. Li, C. Liu, The size effect on void growth in ductile materials, J. Mech. Phys. Solids2. 51 (2003) 1171–1187.

[54] S.J. Moss, G.A. Webster, E. Fleury, Creep deformation and crack growth behavior of a single crystal nickel base superalloy, Metall. Mater. Trans. A. 27 (1996) 829–837.

[55] J. Komenda, P.J. Henderson, Growth of pores during the creep of a single crystal nickel-base superalloy, Scr. Mater. 37 (1997) 1821–1826.

[56] E. Maire, O. Bouaziz, M. Di Michiel, C. Verdu, Initiation and growth of damage in a dual-phase steel observed by X-ray microtomography, Acta Mater. 56 (2008) 4954–4964. doi:https://doi.org/10.1016/j.actamat.2008.06.015.

[57] A. Weck, D.S. Wilkinson, E. Maire, H. Toda, Visualization by X-ray tomography of void growth and coalescence leading to fracture in model materials, Acta Mater. 56 (2008) 2919–2928. doi:https://doi.org/10.1016/j.actamat.2008.02.027.

[58] C. Landron, E. Maire, O. Bouaziz, J. Adrien, L. Lecarme, A. Bareggi, Validation of void growth models using X-ray microtomography characterization of damage in dual phase steels, Acta Mater. 59 (2011) 7564–7573.

[59] Y. Bai, X. Teng, T. Wierzbicki, On the application of stress triaxiality formula for plane strain fracture testing, J. Eng. Mater. Technol. 131 (2009) 1–10.

[60] K.M. Knowles, P.R. Howie, The directional dependence of elastic stiffness and compliance shear coefficients and shear moduli in cubic materials, J. Elast. 120 (2015) 87–108.

[61] S. Ghorbanpoura, M. Zecevic, A. Kumar, M. Jahed, J. Bicknell, L. Jorgensen, I.J. Beyerlein, M. Knezevic, A crystal plasticity model incorporating the effects of precipitates in superalloys: Application to tensile, compressive, and cyclic deformation of Inconel 718, Int. J. Plast. 99 (2017) 162–185.

[62] A.J. Wilkinson, D. Randman, Determination of elastic strain fields and geometrically necessary dislocation distribution near nanoindents using electron back scatter diffraction, Philos. Mag. 90 (2010) 1159–1177.

[63] M.J. Nemcko, J. Li, D.S. Wilkinson, Effects of void band orientation and

crystallographic anisotropy on void growth and coalescence, J. Mech. Phys. Solids. 95 (2016) 270–283.

[64] Y. Alinaghian, M. Asadi, A. Weck, Effect of pre-strain and work hardening rate on void growth and coalescence in AA5052, Int. J. Plast. 53 (2014) 193–205.

[65] Y. Guo, T.L. Burnett, S.A. McDonald, M. Daly, A.H. Sherry, P.J. Withers, 4D imaging of void nucleation, growth, and coalescence from large and small inclusions in steel, SSRN. (2019).

[66] R. Watanabe, Possible slip systems in body centered cubic iron, Mater. Trans. 47 (2006) 1886–1889.

[67] X. Xiong, D. Quan, P. Dai, Z. Wang, Q. Zhang, Z. Yue, Tensile behavior of nickel-base single-crystal superalloy DD6, Mater. Sci. Eng. A. 636 (2015) 608–612.

[68] B.S. El-Dasher, B.L. Adams, A.D. Rollett, Viewpoint: experimental recomvery of geometrically necessary dislocation density in polycrystals, Scr. Mater. 48 (2003) 141–145.

[69] J. Berry, N. Provatas, J. Rottler, C.W. Sinclair, Phase field crystal modelling as a unified atomistic approach to defect dynamics, Phys. Rev. B. 89 (2014) 214117.

[70] V. Navarro, O. Rodriguez de la Fuente, A. Mascaraque, J.M. Rojo, Uncommon dislocation processes at the incipient plasticity of stepped gold surfaces, Phys. Rev. Lett. 100 (2008) 105504.

[71] L.D. Nguyen, D.H. Warner, Improbability of Void Growth in Aluminum via Dislocation Nulceation under Typical Laboratory Conditions, Phys. Rev. Lett. 108 (2012) 035501.

[72] B. Chen, J. Jiang, F.P.E. Dunne, Is stored energy density the primary meso-scale mechanistic driver for fatigue crack nucleation?, Int. J. Plast. 101 (2018) 213–229.